\documentclass[aps,prd,preprint,a4paper]{revtex4}
\usepackage{epsfig}
\begin{document}
\title{ON THE BROKEN GAUGE, CONFORMAL AND DISCRETE SYMMETRIES
IN PARTICLE PHYSICS}       
\author{
D. PALLE \\Department of Theoretical Physics, Rugjer Bo\v skovi\'c 
Institute, \\P.O.Box 1016, Zagreb, CROATIA}
%\date{ }
\maketitle
{\bf Summary.}$-$ Relationships between gauge, conformal and discrete
 symmetries in particle physics are analysed.
 We study also the effect of the electroweak mixing on the
cancellation of SU(2) anomalous actions. It is shown that the relation
 $\theta_{W}=2(\theta_{12}+\theta_{23}+\theta_{13})$
between the Weinberg angle and the Cabibbo-Kobayashi-Maskawa angles should
be satisfied and this effect is completely defined by the mixing of Dirac
fermions. We compare two mechanisms of the spontaneous breaking of gauge
symmetry, discuss the renormalizability of theories, 
and argue for the existence of the Majorana fermions necessary to
remove the SU(2) anomalous action. The fate of the majoron and the spontaneously
broken lepton number is discussed. We also show the compatibility of the
boson and fermion mixings with Dyson-Schwinger equations. \\
 \\
\begin{eqnarray*}
PACS\ numbers:&&11.15.Ex\ Spontaneous\ breaking\ of\ gauge\ symmetry   \\
&&11.30.Er\ Charge\ conjugation,\ parity,\ time\ reversal,\\&&and\ other
\ discrete\ symmetries \\
&&12.10.Dm\ Unified\ theories\ and\ models\ of\ strong\\&& and\ electroweak
\ interactions
\end{eqnarray*}  
        
\vspace{15 mm}

"393. ... Nec Geometrica argumenta quidquam probant in mea Theoria pro
divisibilitate ultra eum limitem; posteaquam enim deventum fuerit
ad intervalla minora, quam sit distantia duorum punctorum, sectiones
ulteriores secabunt intervalla ipsa vacua, non materiam." \\
P. Rogerio Josepho Boscovich Societatis Jesu,  Theoria Philosophiae Naturalis,
Pars Tertia, Venetiis MDCCLXIII. \\

\vspace{7 mm}

\section{Introduction}      
                
In this paper we want to reanalyse gauge theories in such a way as to try to
understand relations between gauge, conformal and discrete symmetry
breaking observed in particle physics.
Two questions arise very naturally: (1) Is it a coincidence that discrete
symmetry breaking happens together with the SU(2) gauge symmetry
breaking? (2) Can we explain the breaking of essentially spacetime symmetries,
such as conformal and discrete symmetries in particle physics, by some
characteristic of the physical spacetime and relate it to the gauge
symmetry breaking?
We try to answer to these questions, studying nonperturbative
anomalies, conformal symmetry and bootstrap equations.

      In the analysis of gauge quantum field theories
   (QFT) anomalies play a crucial role setting the consistency requirements
    on the theory. Calculating quantum fermion loops with 
   pseudoscalar or axial-vector couplings, one can account for all perturbative
   anomalies. The axial anomaly ( Adler-Bell-Jackiw anomaly)\cite{ABJ}  
   was first
   discovered within the effective chiral theories and it could 
   account for the nonvanishing $\pi_{0}\rightarrow 2\gamma$
   amplitude. On the contrary, the Adler-Bardeen anomaly \cite{AB},
    found as an anomalous part of the Ward-Takahashi
   identities, should vanish in order to preserve the renormalizability of the 
   theory. In the standard electroweak model\cite{GSW}, the anomalous part comes
 from the axial-vector coupling and it was shown that higher loops neither   
   affected nor altered the form of the anomaly. The anomalous terms could
be removed by satisfying
   the constraint on the sum of hypercharge assignments of weak
   doublets in the model.

    In this paper, however, we want to analyze essentially nonperturbative
   anomalies arising from quantum fluctuations in the generating functional.
                              In Sec. 2
    we try to understand the source of this type of anomalies
   and how to remove the unwanted terms. In Sec. 3 we 
   discuss the role of zero-mode fermion eigenstates in the cancellation
of anomalous actions, resulting in new unexpected results peculiar
to the SU(2) symmetry. Section 4 is devoted to the study of the 
 quantization and renormalizability 
of the Higgs- and non-Higgs-type spontaneously broken gauge theories.
In Section 5 we study trace anomaly, whereas in Section 6 
 we resolve the problem of the spontaneously broken lepton
number accompanied with the existence or nonexistence of the majoron.
Finally, the electroweak mixing is introduced nonperturbatively in 
the generating functional for the electroweak theory (Sec. 7) and 
the compatibility of the angle relation with 
Dyson-Schwinger equations is proved.
 The concluding section presents a comparison of the 
   phenomenology with theoretical constraints and predictions, and
some general remarks.

\section{SU(2) global anomaly}      
           
  Perturbative anomalies of the Green functions test the 
   local gauge symmetry of the QFT, whereas nonperturbative anomalies test
   the global gauge symmetry through actions and 
   generating functionals.   
   Since the central point in the gauge QFT is always the problem of 
   symmetry breaking, the possible impact of nonperturbative anomalies
on the model building and symmetry-breaking mechanisms might be very important.
      
      The first nonperturbative study of anomalies was performed by
   Fujikawa \cite{Fuj}. He also realized the necessity of referring 
   to the index theorems \cite{At-Si} 
   of the Weyl and Dirac operators. Thus one is then immediately faced with the 
   Wess-Zumino actions introduced earlier \cite{W-Z}  into the study of 
   anomalies, and 
   with the role of homotopy pointed out by Witten \cite{Witt} .

      We now repeat some relevant results. To apply mathematical
   theorems, considerations are carried out in the Euclidean space with
   possible compactification to a sphere or a disc. One of the central
   achievements in the physical cognition is the following statement: the
   index of the Dirac operator in D+2 dimensions is equal to the index of
   the Weyl operator in D dimensions\cite{Zum,Alv}. The gauge-orbit parameters 
   perfectly
   fit into the scheme as two additional dimensions. If the difference between
   the number of left- and right-handed zero modes does not vanish, it 
   will cause the nonvanishing of the Wess-Zumino term. However, the meaning of
   this anomalous term is physicaly clear: if the global gauge symmetry is
  broken, this term cannot vanish. We can now raise a few questions: (a) What is
   the relationship between the anomalous terms of the fermion and boson 
   operators? (b) Is there any problem with unitarity when the anomalous
   actions appear in the theory? (c) What is the role of homotopy in the
   study of the nonperturbative anomaly?

   Similarly to the construction of
      the fermion Lagrangian density in the six-dimensional
   space, one can build the gauge boson density which is cubic in 
   the field strength \cite{Zum,Fadd}. 
   The calculus of quantum fluctuations in the fermion and
   boson determinants then reduces to completely geometrical\cite{Zum}
   calculations using Cartan's forms, cohomology, and Stokes' theorem. 
   However, we should be very cautious about the relative sign between two
   anomalous actions because the integration over the Grassmannian (fermionic,
   anticommutative) variables yields an additional negative sign.

      Gauge non-invariant Wess-Zumino actions are nonhermitian 
   structures that should
 disappear because they spoil the unitarity at the quantum-loop level
\cite{Alv,Fadd}.
   The most natural way of solving this problem is to choose the 
   same gauge fields in the boson and fermion part of the action.
 Then the cancellation of the Wess-Zumino actions will be automatic owing to 
   the additional sign obtained after the Grassmannian integration:

\begin{eqnarray}
ind({\cal F}^{3})=-ind(\not{\cal D}_{6})=
-ind(\not{\partial}_{4}+\not{A}\frac{1}{2}(1-{\gamma}_{5})), \\
ind(\not{\cal D}_{6})=\frac{i^{3}}{{(2\pi)}^{3}3!}\int_{S^{5}}
[\omega_{5}(A^{g},F^{g})-\omega_{5}(A,F)],\nonumber \\
\omega_{5}(A^{g},F^{g})=\omega_{5}(A,F)+d\alpha_{4}+\frac{1}{10}
Tr{(g^{-1}dg)}^{5}, \nonumber
\end{eqnarray}
where $F=dA+A^{2}$ , $g\epsilon SU(2)$ on the boundary of $S^{5}$.

      The most detailed information about the topology comes from the homotopy
   groups. For considerations in the gauge QFT, we need to know the homotopy
   groups and their sequences of unitary symmetries in the four- and higher-
dimensional spaces. Let us recall, for example, the following homotopy 
   sequence\cite{Stee}:
\begin{eqnarray}
\pi_{5}(SU(3))\stackrel{j}{\rightarrow}\pi_{5}(SU(3)/SU(2))
\stackrel{\partial}{\rightarrow}\pi_{4}(SU(2))\stackrel{i}{\rightarrow}
\pi_{4}(SU(3)),\\
j=inclusion\ map,\ i=group\ homomorphism, \nonumber\\
\ \ \partial=boundary\ homomorphism.\nonumber
\end{eqnarray}                                                                           
       This sequence shows that the SU(2) subsymmetry of the 
   SU(3) symmetry will be broken, because the topological defects in the two-
dimensional gauge orbit space(zero modes) affect the SU(2) subsymmetry only.
We have chosen this example because in the standard model we are faced with the broken
   SU(2) symmetry\cite{GSW}. If we start with the SU(3) family\cite{Pa} in the six-
dimensional space\cite{Zum,Alv},
    where gauge fields have six components, we have sufficient
   gauge freedom to develop the structure of the whole standard model.
   Evidently, we can obtain:\{$A_{c}^{a}$;c=0,1,2,3,5,6;a=1,...,8;number
   of the gauge degrees of freedom=6$\times$8=48\} 
   and in the standard model in the 
   Minkowski space$\{U(1):4\times1;SU(2):4\times3;SU(3):4\times8$
   ;SU(3)$\times$SU(2)$\times$U(1):4+12+32=48\}.
   There is presently at our disposal\cite{Ban}
    a mechanism that can generate the local gauge theory of
   subsymmetry, called the mechanism of hidden local symmetries. Whereas a
 detailed analysis of this issue can be found elsewhere\cite{Ban}, here we 
point out the necessary ingredients:(a) the local symmetry and subsymmetries of
the basic group G are generated as hidden local symmetries, but at the same time
   preserving the global G symmetry, (b) the interaction of fermions with gauge
   bosons proceeds in the standard way, (c) the appearance of 
   fermion and gauge boson masses just breaks the global symmetry drawn from
   the homotopy sequence. By choosing G=SU(3) as the basic group, one is able
  to recover the standard model at the tree level, because the embedding of the
   SU(2)$\times$U(1) in SU(3) with the fermion representations (color singlet
 and triplet) exists and it is unique\cite{Dra}; the
SU(3) global symmetry is preserved as it should be in QCD\cite{Frit};
if the SU(2) symmetry
   is broken, fermions are coupled chirally asymmetrically to SU(2) bosons
\cite{GSW,Dra}:

\begin{eqnarray}
SU(3) : H_{1} , H_{2}\ generators\ commuting\ with\ all\ others,
\nonumber\\
E(\vec{\mu}) :\ \vec{\mu}=\vec{\alpha},-\vec{\alpha},\vec{\beta},
-\vec{\beta},\vec{\gamma},-\vec{\gamma}\ (\vec{\mu}=root\ vectors),
\nonumber\\
\vec{\alpha}=\frac{1}{\sqrt{3}}\vec{e}_{1},\ \vec{\beta}=\frac{1}
{2\sqrt{3}}\vec{e}_{1}+\frac{1}{2}\vec{e}_{2},\ \vec{\gamma}=
-\frac{1}{2\sqrt{3}}\vec{e}_{1}+\frac{1}{2}\vec{e}_{2},
\end{eqnarray}                         
SU(2)$\times$U(1) embedding :
\begin{equation}
SU(2) :\ \sqrt{3}H_{1},\ \sqrt{6}E(\pm\vec{\alpha});\ U(1) : H_{2}.
\end{equation}

      Thus we have a perfect unification scheme that has no problem with
   unitarity, and it is not a surprise that it is realized in nature. 
It is worthwhile to notice that the six-dimensional conformal space
\cite{Pa} fits perfectly as Minkowski space plus gauge-orbit two-dimensional
space \cite{Di}.

\section{The SU(2) zero modes}

In this section we show that for the SU(2) symmetry, the Dirac
zero mode fermion eigenstates cannot contribute to the index of the Weyl
operator. To prove this, we closely follow the discussion in Ref.
 \cite{Alv} where a relation is established between the index of
$i{\not D}_{2n+2}$ and the winding number of the phase of the 2n-dimensional
Weyl operator $i{\not D}^{t,\theta}_{+}=i{\not D}^{t,\theta}_{2n}\frac{1}{2}
(1-\gamma_{5})$.
Instead of the operator $i{\not D}_{2n+2}$, the authors of Ref.\cite{Alv} 
studied the following deformed operator with
the same zero mode space:

\begin{eqnarray}
i{\not D}^{\epsilon}_{2n+2}=\frac{1}{\epsilon}i\sum^{2n}_{\mu=1} D_{\mu}
\Gamma^{\mu}+i{\not D}_{2},\nonumber
\end{eqnarray}
  
as a matter of fact, the square of this operator:

\begin{eqnarray}
H_{\epsilon}&\equiv&(i{\not D}^{\epsilon}_{2n+2})^{2}\nonumber\\
&=&\frac{1}{\epsilon^{2}}(iD_{\mu}\Gamma^{\mu})^{2}+
(iD_{i}\Gamma^{i})^{2}+\frac{i^{2}}{\epsilon}\Gamma^{i}\Gamma^{\mu}
(D_{i}D_{\mu}-D_{\mu}D_{i}).\nonumber
\end{eqnarray}
             
The zero modes with a definite chirality ($\chi$) take the form

\begin{eqnarray}
\chi=+1:\ \left(\begin{array}{c}\psi\\0\end{array}\right)\ ;\ \chi=-1:
\ \left(\begin{array}{c}0\\\psi\end{array}\right)\nonumber .
\end{eqnarray} 

The locally Euclidean coordinates $\phi_{1},\phi_{2}$ are introduced in the vicinity of
each zero eigenvalue of the operator $i{\not D}^{t,\theta}_{2n}$ and the space of
zero modes is spanned by two states: 

\begin{eqnarray}
\psi_{\pm}(x)\ \equiv\ \frac{1}{2}(1\mp\gamma_{5})\psi^{\phi_{i}=0}(x).
\nonumber
\end{eqnarray}

We need to investigate an operator in the vicinity of the zero mode,
 so its perturbation near $\phi_{i}=0$ is of the form

\begin{eqnarray}
\delta(i{\not D}^{\phi_{1},\phi_{2}}_{2n})=\sum_{j}(i\partial_{j}{\not A})\phi_
{j},\nonumber\\
where\ \partial_{j}{\not A}\equiv(\partial/\partial\phi_{j}){\not A}^{\phi_{1},
\phi_{2}}\mid_{\phi_{i}=0},\nonumber
\end{eqnarray}

and its matrix elements in the ($\psi_{+},\psi_{-}$) basis are

\begin{equation}
\begin{array}{c}
\left(\begin{array}{cr}0 &\sum_{i}z^{*}_{i}\phi_{i}\\\sum_{i}z_{i}\phi_{i}&0
\end{array}\right),
\end{array}
where\ z_{i}\equiv(\psi^{\dagger}_{-},i\partial_{i}{\not A}\psi_{+}),
\ are\ complex\ constants.\nonumber
\end{equation}

Then in the vicinity of $\phi_{i}=0$ two zero modes have eigenvalues:

\begin{eqnarray}
i{\not D}^{\phi_{i}}_{2n}\Psi^{\phi_{i}}_{\lambda}(x)=
\lambda(\phi_{i})\Psi^{\phi_{i}}_{\lambda}(x),\nonumber\\
\lambda(\phi_{i})=\pm{\mid}z_{1}\phi_{1}+z_{2}\phi_{2}\mid.\nonumber
\end{eqnarray}

After further considerations the authors of Ref.\cite{Alv} obtained
an expression
for the chirality of the zero mode:

\begin{eqnarray}
\chi\ =\ \frac{{\mid}z_{1}\mid{\mid}z_{2}\mid}{Im(z^{*}_{1}z_{2})}.
\end{eqnarray}

Now let us repeat how Dirac and Majorana states are formed from 
two-component Weyl spinors. For a $2\times2$ unimodular matrix $L$
belonging to SL(2,C), Weyl spinors transform under Lorentz transformation
as

\begin{eqnarray}
{\eta}_{a}\ \rightarrow\ {\eta}_{a}'=L^{b}_{a}{\eta}_{b},\nonumber\\
\dot{\eta}_{\dot{a}}\ \rightarrow\ \dot{\eta}_{\dot{a}}'=
{L^{*}}^{\dot{b}}_{\dot{a}}\dot{\eta}_{\dot{b}}\nonumber.
\end{eqnarray}

It is important to stress that $L$ and $L^{*}$ are not equivalent representations
of SL(2,C) \cite{Rue}. A four-component Dirac spinor can now be formed from two Weyl
spinors:

\begin{eqnarray}
\Psi_{D}\ =\ \left(\begin{array}{c}\eta_{a}\\\dot{\xi}^{\dot{a}}\end{array}
\right),\nonumber
\end{eqnarray}

and its charge-conjugate spinor is

\begin{eqnarray}
\Psi^{c}_{D}\ =\ C\bar{\Psi}^{T}_{D}\ =\ \left(\begin{array}{c}\xi_{a}\\
\dot{\eta}^{\dot{a}}\end{array}\right).\nonumber
\end{eqnarray}

Majorana states are defined by the Majorana condition:\newline
\vspace{4 mm}
$\begin{array}{c}
\Psi_{M}\ =\ \left(\begin{array}{c}\eta_{a}\\\dot{\eta}^{\dot{a}}\end{array}
\right),\ \left(\Psi_{M}\right)^{c}=C\left(\bar{\Psi}_{M}\right)^{T}
=\Psi_{M}\ ,\\
C\ =\ \left[\begin{array}{cr}\epsilon_{ab}&0\\0&\epsilon^{\dot{a}\dot{b}}
\end{array}\right],\ \eta^{a}=\epsilon^{ab}\eta_{b},
\ \dot{\eta}^{\dot{a}}=\epsilon^{\dot{a}\dot{b}}\dot{\eta}_{\dot{b}},\\
\epsilon^{12}=1,\ \epsilon_{12}=-1.
\end{array}$
\vspace{4 mm}
\newline
Let us now see chiralities of the SU(2) zero modes. The calculation of
the index takes into account the summation over the chiralities of all 
independent zero modes. If a Dirac state is a zero mode of some gauge
symmmetry, then its charge-conjugate state is not necessarily a zero 
mode. However, the charge-conjugate SU(2) Weyl states are equivalent.
This is peculiar to the SU(2) group\cite{Rue}, and this equivalence could be easily
understood because the generators $\tau^{i}$ and its complex-conjugates $\tau^{i*}$
are connected through the similarity transformation. As a consequence,
a charge-conjugate SU(2) Dirac spinor is also a zero mode if its
appropriate Dirac spinor is a zero mode:

\begin{eqnarray}
\Psi_{D}\equiv\left(\begin{array}{c}\eta_{a}\\\dot{\xi}^{\dot{a}}\end{array}
\right),\ 
\Psi^{c}_{D}=\left(\begin{array}{c}\xi_{a}\\\dot{\eta}^{\dot{a}}\end{array}
\right)\ \stackrel{\rm equiv.}{\simeq}\ \left(\begin{array}{c}\dot{\xi}_
{\dot{a}}\\\eta^{a}\end{array}\right).\nonumber
\end{eqnarray}

In the $(\psi_{+},\psi_{-})$ basis, the perturbation of the $i{\not D}^{\phi
_{1},\phi_{2}}_{2n}$ 
operator for the charge-conjugate Dirac spinor is of the form

\begin{equation}
\hspace{30 mm}\left(\begin{array}{cr}0&\sum_{i}z_{i}\phi_{i}\\\sum_{i}z^{*}_{i}\phi_{i}&0
\end{array}\right).
\end{equation}

From the formula for the chirality of the zero mode it follows:

\begin{eqnarray}
\chi(\Psi_{D})=\frac{\mid{z_{1}}\mid\mid{z_{2}}\mid}{Im(z^{*}_{1}z_{2})}
=-\frac{\mid{z_{1}}\mid\mid{z_{2}}\mid}{Im(z_{1}z^{*}_{2})}=
-\chi(\Psi^{c}_{D}).
\end{eqnarray}

Thus the total contribution of the SU(2) Dirac zero modes always vanishes.
On the contrary, a Majorana state by definition does not have an independent
charge-conjugate spinor, but it has a definite chirality.
The index theorem and the necessity of the exact cancellation of
SU(2) anomalous actions (Sec. 2) forces the existence of the surplus 
of the left-handed (positive chirality) zero modes.

\section{Quantization of spontaneously broken
gauge theories}

The path integral or the canonical quantization of spontaneously broken
non-Abelian gauge theories have been consistently carried out in the past
quarter of the century. However, in the light of the results obtained in the
preceding section, we also want to include into consideration 
Majorana particles. The second reason for reconsideration lies
in our general unsatisfaction with the Higgs mechanism, as an example
of the spontaneously broken gauge theory realized in the electroweak
sector of the Standard Model(SM). 
             
Let us write the most general form of the electroweak Lagrangian with 
leptons, together with the gauge-fixing and the Faddeev-Popov(FP) terms,
which should be included for the correct quantization of the
spontaneously broken non-Abelian gauge theory\cite{Kugo,Aoki}:

\begin{equation}
\begin{array}{c}
{\cal L}\ =\ {\cal L}_{lep}+{\cal L}_{g.bos}+{\cal L}_{scal}+
{\cal L}^{D}_{Yuk}+{\cal L}^{M}_{Yuk}+{\cal L}_{g.fix}+{\cal L}_{FP},
\end{array}
\end{equation}
$
\begin{array}{l}
{\cal L}_{lep}=\bar{R}i\gamma^{\mu}(\partial_{\mu}+ig'B_{\mu})R+
\bar{L}i\gamma^{\mu}(\partial_{\mu}+\frac{i}{2}g'B_{\mu}-
ig\frac{\tau^{i}}{2}A^{i}_{\mu})L+\bar{\psi}_{R}i\gamma^{\mu}\partial_{\mu}
\psi_{R}, \nonumber\\
{\cal L}_{g.bos}=-\frac{1}{4}F^{i}_{\mu\nu}F^{i\mu\nu}-\frac{1}{4}
B_{\mu\nu}B^{\mu\nu}, \nonumber\\
{\cal L}_{scal}=(\partial_{\mu}\Phi^{\dagger}+i\frac{g'}{2}B_{\mu}\Phi
^{\dagger}+i\frac{g}{2}\tau^{i}A^{i}_{\mu}\Phi^{\dagger})(\partial^{\mu}
\Phi-i\frac{g'}{2}B^{\mu}\Phi-i\frac{g}{2}\tau^{i}A^{i\mu}\Phi)-
V(\Phi), \nonumber\\
{\cal L}_{Yuk}^{D}=-Y^{e}_{D}\bar{L}\Phi{R}-Y^{\psi}_{D}\bar{L}\tilde{\Phi}
\psi_{R}+h.c., \nonumber\\
{\cal L}_{Yuk}^{M}=-Y^{\psi}_{M}\bar{L}\tilde{\Phi}(\psi^{c})_{R}+h.c.,
\nonumber\\
definitions:\ e^{-}=charged\ lepton,\ \psi=neutral\ Dirac\ lepton,\nonumber\\ 
e_{L,R}^{-}=left(right)\ chiral\ projection;\nonumber\\
L=\left(\begin{array}{c}\psi_{L}\\e^{-}_{L}\end{array}\right),\ R=e_{R}^{-};\nonumber\\
F^{i}_{\mu\nu}=\partial_{\mu}A^{i}_{\nu}-\partial_{\nu}A^{i}_{\mu}+
g\epsilon^{ijk}A^{j}_{\mu}A^{k}_{\nu},\ B_{\mu\nu}=\partial_{\mu}B_{\nu}-
\partial_{\nu}B_{\mu};\nonumber\\
SU(2)_{L}\ scalar\ doublets:\ \Phi=\left(\begin{array}{c}\phi^{+}\\{\phi^{0}}
\end{array}\right),\ \tilde{\Phi}=i\tau^{2}\Phi^{*}=\left(\begin{array}{c}
\phi^{0*}\\-\phi^{-}\end{array}\right),\nonumber\\
\phi^{0}(x)\ =\ (v+H(x)+i\chi^{0}(x))/{\sqrt{2}},\ v=symmetry\ breaking\ 
parameter,\nonumber\\H=physical\ Higgs\ scalar,\ \phi^{\pm},\chi^{0}=Nambu-
Goldstone\ scalars.\nonumber
\end{array}
$

The Majorana mass terms for charged fermions are forbidden by charge
conservation. If the scalar doublets carry no lepton number, then
the Majorana mass term is excluded, provided that only the spontaneously broken lepton
number is allowed. This is the situation in the SM where only Dirac fermions
receive masses through the Higgs mechanism. However, in this case
it is not possible to have a Majorana zero mode to break the $SU(2)_{L}$
gauge symmetry at the tree level by the mass terms. As a consequence, the
SM seems to be anomalous with respect to the nonperturbative $SU(2)$
anomaly. Of course, it does not affect the perturbative unitarity and
renormalizability of the model. 

On the other hand, if the scalar doublets carry the lepton number\cite{Pec}

\begin{eqnarray}
\tilde{\Phi}\ \rightarrow\ e^{2i\alpha}\tilde{\Phi},
\ L\ \rightarrow\ e^{i\alpha}L,\nonumber\\
R\ \rightarrow\ e^{i\alpha}R,\ \psi\ \rightarrow\ e^{i\alpha}\psi,
\nonumber
\end{eqnarray} 

then its
interactions with Dirac fermions are forbidden and no tree-level Dirac
masses can be generated. On the contrary, the Majorana masses are generated,
but the lepton number as the global $U(1)$ symmetry of the 
Lagrangian is also
spontaneously broken in addition to the local $SU(2)_{L}$ gauge symmetry.

The fate of the Nambu-Goldstone particle (majoron) will be discussed
in Section 6.
The theory is now unitary with respect to the nonperturbative $SU(2)$
anomaly. Even the Majorana masses could be generated through the Higgs
mechanism, but in this paper we want to refer also to the mechanism proposed
in Ref.\cite{Pa}. 
Namely, it is possible to generate masses to the $SU(2)$
bosons and the tree-level mass to the fermion by analyzing the trace anomaly
under the hypothesis of noncontractible space (see Sec. 5)\cite{Pa}:

\begin{eqnarray}
M_{W}\ =\ M_{Z}\ =\ \frac{\sqrt{6}}{\pi}\frac{g}{2}\frac{\hbar}{cd},\\
\hbar\ =\ \frac{h}{2\pi},\ d\ \simeq\ 6\times10^{-17}\ cm,\nonumber\\
m_{fermion}(tree\ level)\ =\ \frac{3}{2}\frac{\hbar}{cd}\ = 
\ \frac{3}{2}\Lambda\ =\ m_{0}.
\end{eqnarray} 

Now, the question is whether this mechanism represents a mechanism of 
spontaneous gauge symmetry breaking. Very general results of Kugo and 
Ojima \cite{Kugo} claim that massive gauge bosons always require
spontaneously broken global charges in a one-to-one correspondence.
Moreover, it was explicitly proved that not only the Higgs mechanism,
but also the symmetry breaking in the Schwinger
model \cite{Ito} and the Nambu-Jona-Lasinio model \cite{Nak} occured
spontaneously. It could be easily seen that if we put 
the physical Higgs scalar field and the Higgs potential equal to zero,
quantize the theory in the $R_{\xi}$-gauge \cite{Tho}
with inclusion of the symmetry-
breaking parameter $v$, and if we proceed smoothly to the U-gauge, then we
obtain the Lagrangian with degenerate massive bosons and Majorana fermions of
the UV-finite theory for $v=\sqrt{6}\Lambda/{\pi}$. The presence of the Nambu-Goldstone
scalars corresponds to spontaneous symmetry breaking, thus a noncontractible spacetime
as a symmetry breaking mechanism satisfies the Kugo-Ojima theorem (\cite{Kugo},
theorem 6.6).

Let us now discuss the issue of the renormalizability of the theories
with and without the Higgs scalar. 
The necessary condition for the spontaneously broken gauge theory to be
renormalizable is to have dimensionless gauge coupling constants, linear
Yukawa coupling of scalars to fermions and up to quartic self-coupling
of scalars\cite{Corn}. However, it is well known \cite{Aoki} that this is not
sufficient to preserve the renormalizability, namely the sufficient 
condition requires 
various identities between Green functions. If the quantized theory
possesses BRST symmetry\cite{Bec}, then, as a consequence, the generalized
 Ward-Takahashi and 
Slavnov-Taylor identities are valid. 
The analysis of Kugo and Ojima \cite{Kugo} of the Higgs-Kible SU(2) model
shows that even in the model with the physical Higgs scalar, the BRST
transformations for the gauge vector, Faddeev-Popov ghosts, Nakanishi-Laudrup
and Nambu-Goldstone asymptotic fields (formulae (4.34) of Ref. \cite{Kugo})
form a closed set of field transformations without the inclusion of the
physical Higgs scalar. Thus the massive non-Abelian gauge theory is
renormalizable even without the inclusion of the Higgs scalar because
the quantized theory preserves the BRST symmetry. The inclusion of fermions
or the electroweak mixing does not alter the conclusion. In the case of the
Abelian theory, FP ghosts are decoupled from the gauge vector and NG scalar
fields and then we need the Higgs scalar to preserve the renormalizability
(see Ref. \cite{App}).

Thus in Eq.(9) we have displayed the Lagrangian for the four renormalizable
electroweak theories and let us denote them by (AX), (AY), (BX), and (BY)
 with the following notations: (A)=scalar doublet does not carry lepton
number, (B)=scalar doublet carries lepton number$=-2$, (X)=Higgs scalar
and its quartic self-coupling are present, (Y)=absence of the Higgs scalar and
the presence of the hypothesis of noncontractible spacetime. In the
next sections we study the (BY) theory.

\section{Trace anomaly and noncontractible spacetime}

The hypothesis of the fundamental length attracted much attention in 
the past, but with poor success. Yukawa and Heisenberg touched the essential
problems with the ultraviolet singularity and bootstrap equations in QFT,
but it would have been premature to expect resolutions at that time when
gauge QFT had been confirmed and established neither experimentally nor
theoretically.

The fundamental (elementary) length appears as a universal parameter in
nonlocal QFT to make interactions nonlocal through some nonlocal
functionals, as in the work of Efimov\cite{Ef}. This theory does not improve 
QFT
but introduces new arbitrariness owing to the free choice of elementary-
length-dependent
 functionals.

Insisting on local gauge theory, Kadyshevsky et al.\cite{Ka} developed a 
theory positioned in the five-dimensional de Sitter space where the 
fundamental length appeared as the upper bound to the masses, and the fifth
coordinate was fixed by the "maximon" mass. However, it seems very dubious 
to set any upper bound to all the masses (timelike region) by the fundamental 
length (cutoff in spacelike region).

We should mention lattice gauge theories\cite{Wi} with discretized spacetime.
In these theories 
the size of hypercubes is a technical rather than a physical parameter
that theorists want to remove from their nonperturbative calculations.
The doubling of the fermion species on the lattice represents an 
insurmountable task for the treatment of lattice gauge theories.

Contrary to the hypotheses mentioned above, we propose the following\cite{Pa}
: (1)
The physical spacetime is a continuum in the sense that spacetime
variables can be continuously changed, (2) the physical spacetime is
noncontractible in the sense that spacelike intervals have a lower bound,
 the universal nonvanishing minimal distance ($d\neq0$).

The question arising here concerns the correspondence between these
postulates and local relativistic quantum field theory. First, it must be
clear that the concept of field discards the concept of point particle.
We describe the quantum field as a wave-packet system, i.e. a very
nonlocal structure. However, the interaction between the fields could be
local in the sense that only coordinates of the centre of motion of the
fields (particles) coincide. The nonlocal structure of the fields and the 
fields in interaction should be defined by the complete structure of
dressed Green functions (propagators and vertices). 

Contrary to the perturbation theory, the calculation  of the trace anomaly
is essentially a nonperturbative field identity, valid to arbitrary quantum 
loop order. We shall study the matrix elements of the trace of the
energy-momentum tensor. Let us consider massles spin-$\frac{1}{2}$ fermions 
coupled to gauge bosons. A part with the spinor of the trace operator is of
the form\cite{Pa}:

\begin{eqnarray}
\theta^{\mu}_{\mu}
 = -3\imath\overline{\psi}\gamma^{\mu}\partial_{\mu}\psi + 
\frac{3}{2}\imath\partial^{\mu}[\overline{\psi}\gamma_{\mu}\psi] ,
\end{eqnarray}

where we have discarded the terms linear in vector gauge fields because
these cannot contribute to the spinor matrix element of the trace
at the tree level.
 We search for the nonvanishing contribution to the mass term and
only terms with the canonical momenta could contribute. Owing to
translational invariance we can choose the reference point $x_{\mu}=0$
of the trace operator (in the QFT only canonical variables are correlated):

\begin{eqnarray}
\theta^{\mu}_{\mu}(0) = -\frac{3}{2}\imath(\psi^{\dag}(0)\dot{\psi}
(0)-\dot{\psi}^{\dag}(0)\psi(0)) ,
\end{eqnarray}

where $\dot{\psi}(0)=min \triangle\psi(0)/min{\triangle}t$,
$ \triangle\psi(t)=t(\partial/{\partial}t)\psi(t)=-{\imath}t\hat{p}^{0}
\psi(t)$. Taking into account Lorentz invariance and the noncontractibility
of the physical space, it follows that ${\triangle}t \geq d/c$ ($c$=velocity
of light)

\begin{eqnarray}
\theta^{\mu}_{\mu}(0)=-\frac{3}{2}\frac{1}{cd}min[\psi^{\dag}
(t\vec{\hat{p}^{0}}+\stackrel{\leftarrow}{\hat{p}^{0}}t)\psi](0)\nonumber\\
= -\frac{3}{2}\frac{1}{cd}min[\psi^{\dag}(t\vec{\hat{p}^{0}}-
\vec{\hat{p}^{0}}t)\psi](0) .
\end{eqnarray}

Planck principle of quantum action (or Heisenberg uncertainty principle)
leads us further to\cite{Pa}

\begin{eqnarray}
<1 fermion \mid\theta^{\mu}_{\mu}(0)\mid1 fermion> = \frac{3}{2} 
\frac{\hbar}{cd} .
\end{eqnarray}

Otherwise, in local gauge QFT the following relation
is valid for massive fermions coupled to arbitrary gauge fields\cite{Cl}:

\begin{eqnarray}
<1\  fermion\mid\theta^{\mu}_{\mu}(0)\mid1\ fermion>\ =\ m(1+\gamma),
\end{eqnarray}

where $m$ is the physical mass of the fermion and $\gamma$ is the sum of
anomalous dimensions of the fermion bilinear operator with respect to the
coupled vector gauge fields.

Obviously, from (15) and (16) we may conclude that $m_{0}=\frac{3}{2}\hbar/cd$,
because at the tree level $\gamma = 0$ by definition. The finiteness of 
distances, velocities and energies is essential for the derivation of the
mass. The role and the meaning of this tree-level mass we discussed
in Sec. 4.

On the mass-shell matrix elements of $\theta^{\mu}_{\mu}$ (or equivalently
$\partial_{\mu}D^{\mu}$) of spin-$\frac{1}{2}$ fermions simultaneously
measure nonvanishing energy densities and the noncontractibility of space.
Generalizing this consideration to all elementary particles and taking
into account physical dimensions, one would expect

\begin{eqnarray}
<1\ fermion\mid\theta^{\mu}_{\mu}(0)\mid1\ fermion>\ \propto\ 
\frac{h}{cd}\ ,\\
<1\ vector\mid\theta^{\mu}_{\mu}(0)\mid1\ vector>\ \propto\ 
(\frac{h}{cd})^{2}\ .
\end{eqnarray}

Massless gluons can satisfy (18) mainly because of the non-Abelian character
of gauge symmetry, Wick's theorem applied to the terms quartic in
$G^{a}_{\mu}$ and the running coupling. Actually, the following 
nonperturbative relation for the trace is valid \cite{Cl}:

\begin{eqnarray}
\theta^{\mu}_{\mu}& =& \frac{\beta(g)}{2g}G^{a}_{\mu\nu}
G^{a\mu\nu}\ ,\\
where\ \beta(g)&=&\lambda\frac{{\partial}g(\lambda)}{\partial\lambda},\ 
G^{a}_{\mu\nu}=\partial_{\mu}G^{a}_{\nu}-\partial_{\nu}G^{a}_{\mu}+
gf^{abc}G^{b}_{\mu}G^{c}_{\nu}.\nonumber
\end{eqnarray}

  In the preceding chapters on anomalies we have emphasized the necessity
of having massive weak bosons to remove the SU(2) chiral anomaly by the
SU(2) boson anomaly. Proceeding as for gluons, the trace anomaly for
SU(2) bosons is of the form\cite{Pa,Kugo}

\begin{eqnarray}
\theta^{\mu}_{\mu}\ =\ 2F^{i}_{\mu\nu}F^{i\mu\nu},\ i=1,2,3,\ 
F^{i}_{\mu\nu}=\partial_{\mu}A^{i}_{\nu}-\partial_{\nu}A^{i}_{\mu}+
g\epsilon^{ikl}A^{k}_{\mu}A^{l}_{\nu}.
\end{eqnarray}

Only the quartic self-coupling terms in (20) can generate the mass term.
Therefore, applying Wick's theorem to the time-ordered product, one
obtains\cite{Pa}

\begin{eqnarray}
\theta^{\mu}_{\mu}(x)&=&2g^{2}\epsilon^{ijk}\epsilon^{ilm}A^{j}_{\nu}(x)
A^{k}_{\lambda}(x)A^{l\nu}(x)A^{m\lambda}(x)+...\nonumber \\
&=&2g^{2}\epsilon^{ijk}\epsilon^{ilm}\ \sum_{PERM.}:A^{j}_{\nu}(x)
A^{k}_{\lambda}(x):\overline{A^{l\nu}(x)A^{m\lambda}(x)}+...\ ,\nonumber \\
\overline{A^{l\nu}(x)A^{m\lambda}(x)}&=&<T(A^{l\nu}(x)A^{m\lambda}(x))>_{0}=
{\imath}g^{\nu\lambda}D^{c}_{0}(0)\delta^{lm},\nonumber \\
D^{c}_{0}(0)&=&-(2\pi)^{-4}{\int}k^{-2}d^{4}k,\ (Wick's\ contraction).
\end{eqnarray}

From (21) it follows that

\begin{eqnarray}
\theta^{\mu}_{\mu}(x)\ =\ 24g^{2}{\imath}D^{c}_{0}(0):A^{i}_{\mu}
(x)A^{i\mu}(x):+...\ \ .
\end{eqnarray}

After performing Wick's rotation and integration in $D^{c}_{0}(0)$ in
Euclidean space, one obtains\cite{Pa}

\begin{eqnarray}
\theta^{\mu}_{\mu}(x)=\frac{3g^{2}}{2\pi^{2}}\Lambda^{2}
:A^{i}_{\mu}(x)A^{i\mu}(x):+...\ \ .
\end{eqnarray}

where $\Lambda = \hbar/cd$ is the ultraviolet (UV) cut-off in UV finite 
theory.

Comparison with the trace anomaly of massive bosons:

\begin{eqnarray}
\theta^{\mu}_{\mu}(x) = M^{2}_{W}:A^{i}_{\mu}(x)A^{i\mu}(x):+...
\end{eqnarray}

immediately gives\cite{Pa}

\begin{eqnarray}
M_{W}\ &=&\ \frac{\sqrt{6}}{\pi}\frac{g}{2}\frac{\hbar}{cd},\   \\
\hbar\ &=&\ \frac{h}{2\pi},\ d\ \simeq\ 6\times10^{-17}\ cm.\nonumber
\end{eqnarray}

We have successfully generated masses at the tree level, but masses of
weak bosons are degenerate, which is in disagreement with reality:
 $M_{W} \neq M_{Z}$ .

Owing to the absence of self-coupling in Abelian gauge QFT it is 
apparently impossible
to satisfy (18).
However, there is a unique resolution of this problem: SU(2)$\times$U(1)
is a subsymmetry of SU(3) and one can propose mixing in the neutral sector
\cite{GSW}:

\begin{eqnarray}
A_{\mu}\ &=&\ cos\theta_{W}\ B_{\mu}+sin\theta_{W}\ A^{3}_{\mu},\nonumber\\
Z_{\mu}\ &=&\ sin\theta_{W}\ B_{\mu}-cos\theta_{W}\ A^{3}_{\mu},\ 
\ \theta_{W} \neq 0.
\end{eqnarray}

Redefinition of the neutral SU(2) boson and the U(1) boson in (26) makes 
a drastic change in the spectrum and the couplings. Substitution of (26)
into the SU(2)$\times$U(1) lagrangian with weak $g$ and electromagnetic $g'$
couplings recovers the standard Glashow-Salam-Weinberg model and as a
consequence\cite{GSW}

\begin{eqnarray}
cos\ \theta_{W}=M_{W}/M_{Z},\ e=g\ sin\theta_{W}=g'\ cos\theta_{W}.
\end{eqnarray}

Owing to (24) the physical photon field $A_{\mu}$ in (24) fulfils
the condition (18)

\begin{eqnarray}
<1\ photon\mid\theta^{\mu}_{\mu}(0){\mid}1\ photon>=sin^{2}\theta_{W}
M^{2}_{W}\neq0.
\end{eqnarray}

\section{Spontaneously broken lepton number and the majoron}

In this Section we want to show that the fate of the majoron is in complete
analogy with the fate of the axion in QCD\cite{Kugo}. We have formed the electroweak
Lagrangian invariant under the global $U(1)$ transformation of the lepton
number. The Adler-Bell-Jackiw
anomaly of leptons coupled chirally-asymetrically to the external SU(2) fields in 
the triangular graph breaks the lepton number-conservation at the quantum-loop
level\cite{Rer}:

\begin{equation}
\begin{array}{c}
\partial^{\mu}j^{LN}_{\mu}=-\frac{g^{2}N_{wd}}{32\pi^{2}}\epsilon_{\mu\nu
\rho\sigma}F^{\mu\nu{i}}F^{\rho\sigma{i}},\ i=1,2,3,\\
j^{LN}_{\mu}=\bar{L}\gamma_{\mu}L+\bar{R}\gamma_{\mu}R+\ldots,\ 
N_{wd}=number\ of\ lepton\ weak\ doublets.\nonumber
\end{array}
\end{equation}

However, the situation can be improved by redefinition of the lepton-number
vector current to restore the conservation of the current, but spoiling 
the gauge invariance of the current\cite{Kugo}:

\begin{equation}
\begin{array}{c}
J^{\mu}_{LN}=j^{\mu}_{LN}+X^{\mu},\\
X^{\mu}=\frac{g^{2}N_{wd}}{16\pi^{2}}\epsilon^{\mu\nu\rho\sigma}
[A^{i}_{\nu}F^{i}_{\rho\sigma}-\frac{g}{3}\epsilon_{ijk}A^{i}_{\nu}
A^{j}_{\rho}A^{k}_{\sigma}].\nonumber
\end{array}
\end{equation}

Let us calculate the following vacuum expectation value:

\begin{eqnarray}
\partial^{x}_{\mu}\langle{0}\mid{T}(J^{\mu}_{LN}(x)(\overline{\psi}_{L}
(\psi^{c})_{R}+\overline{(\psi^{c})}_{R}\psi_{L})(0))\mid{0}\rangle\nonumber\\
=2\delta^{4}(x)\langle{0}\mid(\overline{\psi_{L}}(\psi^{c})_{R}+
\overline{(\psi^{c})}_{R}\psi_{L})(0)\mid{0}\rangle\neq0.\nonumber
\end{eqnarray}

This matrix element does not vanish because of the spontaneously broken
$SU(2)_{L}$ gauge symmetry through the Majorana mass. Now we apply the Goldstone
theorem which asserts that the spontaneously broken global $U(1)$ symmetry leads
inevitably to the existence
of the Goldstone boson:

\begin{eqnarray}
J^{\mu}_{LN}(x)\stackrel{x_{0}{\rightarrow}+\infty}{\longrightarrow}
\partial^{\mu}\chi^{out}(x)+\ldots;\ \chi=Goldstone\ boson=majoron.
\end{eqnarray}

The central question whether this Nambu-Goldstone particle is physical
or not, is answered within the Kugo-Ojima quartet mechanism\cite{Kugo}, and
a straightforward construction of all members of the quartet gives:

\begin{equation}
\begin{array}{c}
[Q_{B},\chi(x)]=-i\gamma(x);\ Q_{B}=BRST\ charge\ operator,\nonumber\\
{\cal C}^{\mu}(x)\equiv[iQ_{B},J^{\mu}_{LN}(x)]=\frac{g^{2}N_{wd}}
{8\pi^{2}}\epsilon^{\mu\nu\rho\sigma}\partial_{\nu}c^{i}
\partial_{\rho}A^{i}_{\sigma}(x),\nonumber\\
{\cal C}^{\mu}(x)\stackrel{x_{0}\rightarrow\infty}{\longrightarrow}
\partial^{\mu}\gamma(x)+\ldots,\nonumber\\   
similar\ construction\ gives:\ \bar{\cal C}^{\mu}(x)=\frac{g^{2}N_{wd}}
{8\pi^{2}}\epsilon^{\mu\nu\rho\sigma}\partial_{\nu}\bar{c^{i}}\partial_{\rho}
A^{i}_{\sigma}(x);\nonumber\\ {\cal B}^{\mu}(x)=\{Q_{B},\bar{\cal C}^{\mu}(x)\},
\ c^{i},\bar{c^{i}}=Faddeev-Popov\ ghosts,\nonumber\\
\bar{\cal C}^{\mu}(x)\stackrel{x_{0}\rightarrow\infty}{\longrightarrow}
\partial^{\mu}\bar{\gamma}(x)+\ldots, {\cal B}^{\mu}(x)\stackrel{x_{0}\rightarrow
\infty}{\longrightarrow}\partial^{\mu}\beta(x)+\ldots,\nonumber\\
\{Q_{B},\bar{\gamma}(x)\}=\beta(x),
\ \ (\chi,\beta,\gamma,\bar{\gamma})=Goldstone\ quartet.\nonumber
\end{array}
\end{equation}

Since any member of the quartet appears only in the 
zero-norm combination, the majoron is an unphysical particle until the
weak interaction is described by a non-Abelian gauge symmetry.
 
\section{Compatibility of the electroweak mixing with bootstrap
equations}

Absence of fermion and weak boson mixing (mass degeneracy)\cite{Pa,Kugo} should be the only
condition for exact cancellation of the anomalous actions. The electroweak
mixing\cite{GSW} affects the $SU(2)$ global charges and the only appropriate object for
nonperturbative study is the generating functional of the electroweak
theory\cite{Fadd}: 

\begin{eqnarray}
Z=\int[dA_{\mu}][d\psi][d\overline{\psi}]e^
{i\int{\cal L}_{EW}(A_{\mu},\psi,\overline{\psi})d^{4}x}\ .
\end{eqnarray}

One can only ad hoc introduce the mixing into the functional measure
and the Lagrangian density. However, this cannot be done completely arbitrarily.
In fact, the total mixing in the boson sector should exactly cancel the total
mixing in the Dirac fermion sector of the functional measure\cite{Pa}:
\begin{equation}
\begin{array}{l}      
[dA_{\mu}][d\psi][d\overline{\psi}]=[dA_{\mu}]_{W^{\pm}}
[d\psi]_{u,c,t}[d\overline{\psi}]_{\overline{u},\overline{c},\overline{t}} \\
\noalign{\smallskip}\mbox{~~~~~~~~~~}
\times V(\theta_{W})\left(\begin{array}{c}[dA_{\mu}]_{\gamma}\\{[dA_{\mu}]}_{Z}
\end{array}\right)V_{CKM}^{-1}\left(\begin{array}{c}[d\psi]_{d}\\{[d\psi]}_{s}
\\{[d\psi]}_{b}\end{array}\right)V_{CKM}^{-1}\left(\begin{array}{c}
{[d\overline{\psi}]}_{\overline{d}}\\{[d\overline{\psi}]}_{\overline{s}}\\
{[d\overline{\psi}]}_{\overline{b}}\end{array}\right);\\
\end{array}
\end{equation}
\begin{equation}
\begin{array}{l}      
V(\theta_{W})(TV_{CKM}^{-2}S)_{2\times2}=1,\mbox{~~~~~~} T,S=unitary\ matrices; \\
\end{array}
\end{equation}
\begin{equation}
\begin{array}{l}      
TV_{CKM}^{-2}S=\left(\begin{array}{cr} V(-2(\theta_{c}+
\theta_{23}+\theta_{13}))&0\\0&1\end{array}\right),\\ 
\noalign{\smallskip}
V(\theta_{W})=\left(\begin{array}{cr}cos\theta_{W}&
sin\theta_{W}\\-sin\theta_{W}&cos\theta_{W}\end{array}
\right); \\
\end{array}
\end{equation}

\begin{equation}
\Rightarrow\theta_{W}=2(\theta_{c}+\theta_{23}+\theta_{13})\ .
\end{equation}

We propose the existence of three fermion families. The relation between
boson and fermion mixing angles should be valid for both quark and lepton
Dirac species. One cannot distinguish particle from antiparticle fermions
in the measure, and what figures is only the
 total rotation of the Dirac-fermion variables deduced
by a biunitary transformation and reduction to  a 2$\times$2 mixing matrix.

The introduction of rotated fields into the Lagrangian density restores
the structure of the electroweak sector of the SM: (1) It does not affect
neutral currents  \cite{GSW} and brings the CKM matrix into the
charged currents \cite{Cab}, but now with a constrained
sum on the Euler angles (see Eq.(37)) for quarks and leptons.
(2) Rotated electroweak neutral gauge bosons account for masses as in the SM
\cite{GSW}.

If we accept the Higgs mechanism for Dirac particles as in the SM, the theory
contains no Majorana zero modes and so it is anomalous with respect to the
nonperturbative SU(2) anomaly. If we accept the Higgs mechanism for Majorana
particles, it is not anomalous, but it remains to be UV-singular in a
nonperturbative sense. Thus one can only calculate perturbative (radiative)
corrections to mass terms. It is most favorable to choose 
a noncontractible spacetime as a symmetry breaking mechanism \cite{Pa} when
the theory is not only renormalizable, but also
 nonanomalous and nonsingular (BY theory).

The study of UV-finite theory allows nonperturbative calculations.
In fact, we have a well-defined set of equations for elementary
particles: Dyson-Schwinger (DS) equations\cite{Dy} and, for bound states,  
Bethe-Salpeter (BS) equations\cite{Be}. Any of two-, three-, and four-point Green's
functions satisfies some bootstrap equations that are coupled and  
connected with Ward-Takahashi and Slavnov-Taylor \cite{Wa} identities.
Using the UV cut-off, $M_{W}, M_{Z}$, the fine-structure constant $\alpha_{e}$,
 the tree-level fermion mass, the angle constraint (37), and the QCD vertices,
 we should be able to find lepton and quark masses 
 (see Figs. 1 and 2). 

The most important features of bootstrap systems can be summarized
as follows: (1) The structure and mass singularities of fermion 
propagators in the timelike region are defined by the evolution from
the spacelike region \cite{Fu}. (2) It is essential for fermions
to couple to the massive gauge boson in order to develop a fermion mass 
singularity of the propagator
 \cite{Fu,At}. (3) The tree-level fermion mass appears only in the
equations for Majorana particles. (4) The appearance of three fermion
families could only be interpreted as zero-, one-, and two-node solutions
of the DS equations \cite{Fu,LeY}. (5) Mass gaps between fermion
families are in accordance with the behavior of solutions with
different number of nodes, especially with the slopes at $p^{2}=0$ of
the mass functions \cite{LeY}. (6)  The complete mass matrix for neutral 
leptons for the three families is a 6$\times$6 matrix of the form 
\cite{Bi,Gel}

\begin{eqnarray}
(\overline{f},\overline{F})\left[\begin{array}{cr}M_{L}&M_{D}\\
M_{D}&M_{R}\end{array}\right]\left(\begin{array}{c}f\\F\end{array}
\right)\ ,\ \ \ \ \ \ \ \ \ \ \ \ \ \ \nonumber\\
f\ =\ \frac{1}{\sqrt{2}}(\psi_{L}+(\psi_{L})^{c}),\ \ 
F\ =\ \frac{1}{\sqrt{2}}(\psi_{R}+(\psi_{R})^{c})\ .\\
M_{D}\ =\ Dirac\ mass\ matrix,\ M_{L,R}\ =\ Majorana\ mass\ matrices.
\nonumber
\end{eqnarray}

(7)  Because of the absence of the right-handed currents, the Majorana
mass matrix $M_{R}$ vanishes. Only the Dirac sector is responsible for
flavor mixing and the Majorana sector is flavor diagonal. Thus, after
diagonalizing the off-diagonal Dirac mass matrix, we obtain six
Majorana physical states for which the seesaw relation\cite{Gel} is valid for 
each family separately:

\begin{eqnarray}
m_{D}{\ll}m_{N},\ \ m_{\nu}\simeq\frac{m^{2}_{D}}{m_{N}},\nonumber\\
\psi_{N}{\simeq}f-\frac{m_{D}}{m_{N}}F,\ \ \psi_{\nu}\simeq\gamma_{5}
(F+\frac{m_{D}}{m_{N}}f)\ ,\nonumber\\
\psi_{N},\ \psi_{\nu}\ =\ Majorana\ particles.
\end{eqnarray}

(8) The light Majorana particles, namely, neutrinos, can interact with the
left-handed currents because (${\psi}^{c})_{L}=({\psi}_{R})^{c},\ 
\psi^{c}=C\overline{\psi}^{T}$ (one can also envisage this in terms of
Weyl's spinors).
(9) The Dirac masses are much smaller than the Majorana masses $m_{N}$
because of 
 no tree-level mass and smaller effective
couplings ($\theta_{W}=0$ implies $m_{D}=diag(m_{1},m_{2},m_{3})$).
(10) The Dirac masses of the first family are ${\cal O}(0.1MeV)$. However, 
because of 
the strong interaction we cannot measure quark flavour masses and there are no
quark-mass singularities \cite{Mar}(massless gluons prevent their appearance), 
which is interpreted as confinement. In QCD, one uses values of quark  
current mass functions  
which are much higher than flavour-mass eigenstates (see Fig. 2).
 (11) An 
 essential contribution to the heavy Majorana fermion masses of the second
and third family comes from the longitudinal strongly interacting part  
in quantum loops
(Nambu-Goldstone mode), with $g^{2}_{eff}=1/16sin^{2}\theta_{W}\times
\alpha_{e}/\pi(m_{N}/M_{W})^{2},\ m_{N}{\gg}M_{W}$, but the heavy Majorana
lepton of the first family acquires the mass $m_{N_{e}}{\simeq}m_{0}=485\ GeV$.
(12) The electron neutrino has the mass $m_{{\nu}_{e}}={\cal O}(0.1eV)$. 
(13) Considerable progress has recently been made in solving the BS equations
for the quarkonium, based on modeling propagators \cite{Mun}.
(14) In QED, perturbation theory can still be successfully applied to the BS
system for the positronium\cite{Nshi}.
(15) The theory is free of the Adler-Bardeen anomaly\cite{AB} and it contains 
60 elementary particles, 48 fermions, and 12 vectors, so in fact there is
an equal number of spin-$\frac{1}{2}$ and spin-0 species (12 vectors are
equivalent to 12$\times$4=48 scalars)\cite{Pa}

\begin{equation}
\begin{array}{c}      
\left[\ \begin{array}{c} A_{\mu} \\ W^{-}_{\mu} \\ W^{+}_{\mu} \\
 Z_{\mu} \end{array}\right]\ ;\ \left[\ \begin{array}{c} \nu_{l} \\
 l^{-} \\ l^{+} \\ N_{l} \end{array}\right], l=e,\mu,\tau\ ;
\left[\begin{array}{c} \overline{u}^{k}_{f} \\ \overline{d}^{k}_{g} \\
 d^{k}_{g} \\ u^{k}_{f} \end{array}\right],\\
 \ \left(\begin{array}{c}f\\g\end{array}
\right)\ =\ \left(\begin{array}{c}u,c,t\\d,s,b\end{array}\right),
\ \ k=1,2,3;\ \left[G^{a}_{\mu}\right],\ a=1,...,8.\ \ \ \ \ \ 
\end{array}
\end{equation}

\section{Conclusions}

In this paper we have made an attempt
to understand and resolve the problem of the $SU(2)$ nonperturbative anomaly
and its possible connection with the electroweak mixing. The resolution
of the problem does not favour the SM, but the theory with Majorana
leptons. A qualitative study of the spectrum of the UV-finite
theory using
 the gauge symmetry-breaking mechanism based
on the characteristic of the physical spacetime, namely, its noncontractibility,
leads to surprisingly successful results.
It could be plausible that the breaking of gauge, discrete\cite{Lee}, and conformal symmetries
has a common source, i.e. noncontractible spacetime.
 Whereas the weak-interaction scale defines the measure of  
noncontractibility, the conformal space with the SU(3) symmetry describes the complete
vector gauge structure of the SM. A study of UV-finite bootstrap equations
can explain the appearance of fermion families, flavor mixing, small neutrino
masses, mass gaps between fermion families, and mass gaps between Majorana 
 and Dirac fermions.

In the quark sector it is possible to check an angle constraint \cite{Jar,RPP}:

\begin{eqnarray*}
at\ 90\%\ CL\ \ \left\{\begin{array}{l}sin\theta_{c}=0.221\pm0.003
\\sin\theta_{23}=0.040\pm0.008
\\sin\theta_{13}=0.0035\pm0.0015\end{array}\right. \\
at\ 68\%\ CL\ sin^{2}\theta_{W}=0.2319\pm0.0007\\
define\ \ \bigtriangleup\theta_{BF}=\theta_{W}-2(\theta_{c}+\theta_{23}
+\theta_{13}),\\ then\ at\ 99\%\ CL\ \bigtriangleup\theta_{BF}=-0.0405\pm
0.0402\ . 
\end{eqnarray*}
                        
The large error comes from the poor knowledge of the Bethe-Salpeter wave
functions of mesons in the calculation of processes with large
momentum transfer necessary to evaluate the $\theta_{23}$ angle.
Recent calculations lower this angle significantly: heavy quark effective
theory gives $\theta_{23}{\simeq}0.0267$ (Ref. \cite{Heavy}) and
Salpeter solutions $\theta_{23}{\simeq}0.032$ (Ref. \cite{Bonn}).

Measurements of the solar neutrino flux and its observed deficiency
could be explained by the Mikheyev-Smirnov-Wolfenstein mechanism\cite{Mik},
 which requires massive neutrinos with flavor mixing.
The structure-formation analysis with cold and hot dark matter referring to the 
COBE data always contains massive neutrinos. Rotational curves 
and gravitational lensing measurements give strong evidence for dark matter,
and cosmologically stable light and heavy neutrinos are perfect
particle candidates for dark matter \cite{Sci}. Presently, only astronomical and
astrophysical measurements provide some evidence for massive neutrinos, whereas
terrestrial experiments only set bounds on the masses and mixing angles.
However, interesting results may be expected in the near future from new
experiments on the neutrino mixing at CERN(CHORUS, NOMAD), LANL(LSND)
\cite{LSND}, neutrinoless double beta decay, etc.

LEP 1 data could be fairly well fitted by the perturbation theory of the SM\cite{Hol}, but
perturbative calculations could be easily provided in  
the UV-finite theory without Higgs scalars except that instead of regularization
one has to integrate up to the UV-cutoff. The sensitivity of the observables to  the Higgs 
mass or to the UV-cutoff is only logarithmic, so one has to wait for more and better
data from LEP 2. A recent SM model fit of LEP data cannot explain $b$ and $c$
quark channel decay widths of the Z-boson \cite{LEP}. One can also notice a problem to
fit Tevatron data for the cross section for jets at the transverse energies
greater than $200\ GeV$ by perturbative QCD \cite{CDF}.
A definite answer about the source of symmetry breaking will be given by LHC,
which is also capable to detect heavy Majorana fermions up to 700 GeV\cite{ING},
or by some future linear $e^{+}e^{-}$ collider.\\
  \\

\begin{figure}
\begin{center}
\epsfig{file=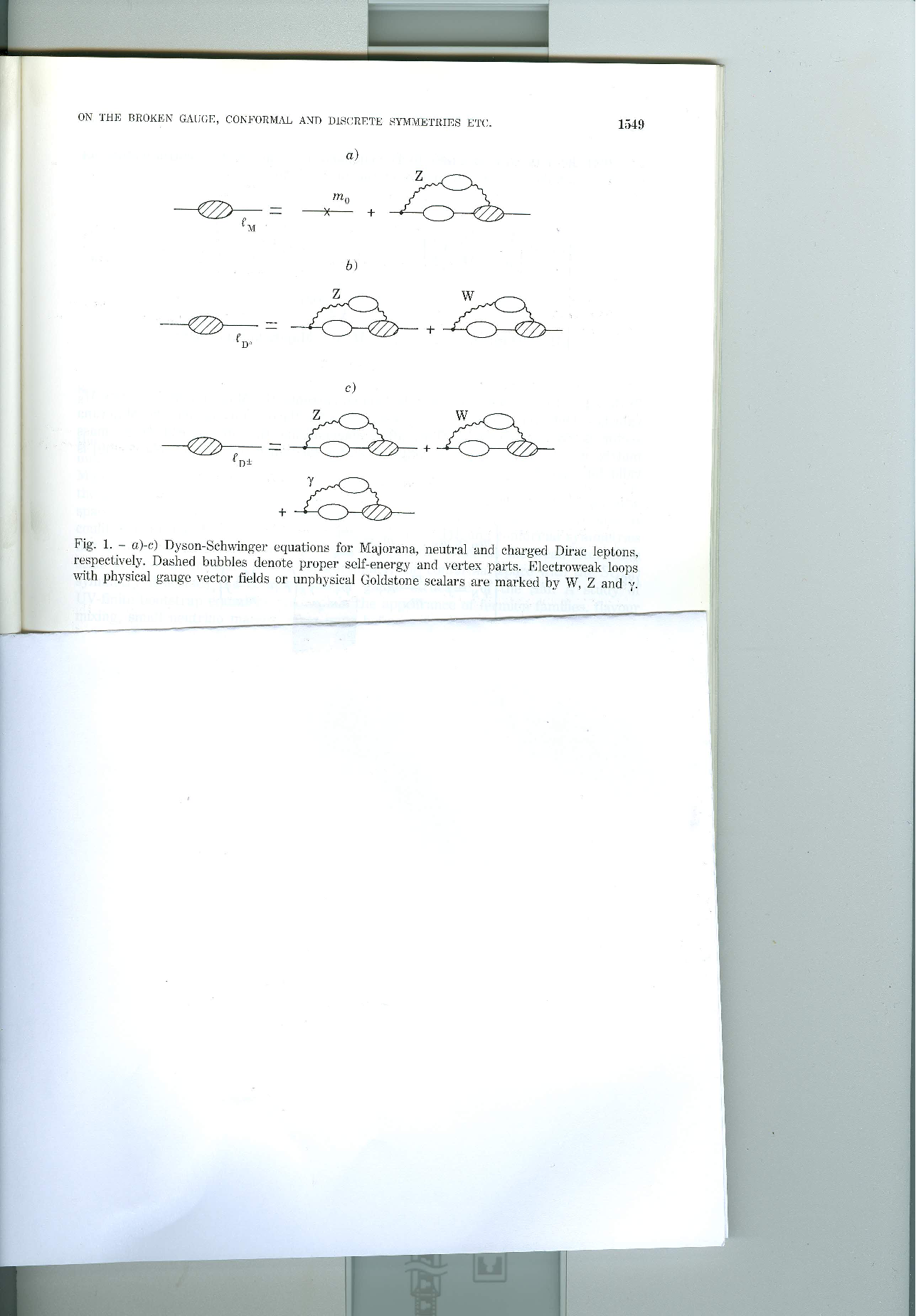}
\end{center}
\end{figure}

\begin{figure}
\begin{center}
\epsfig{file=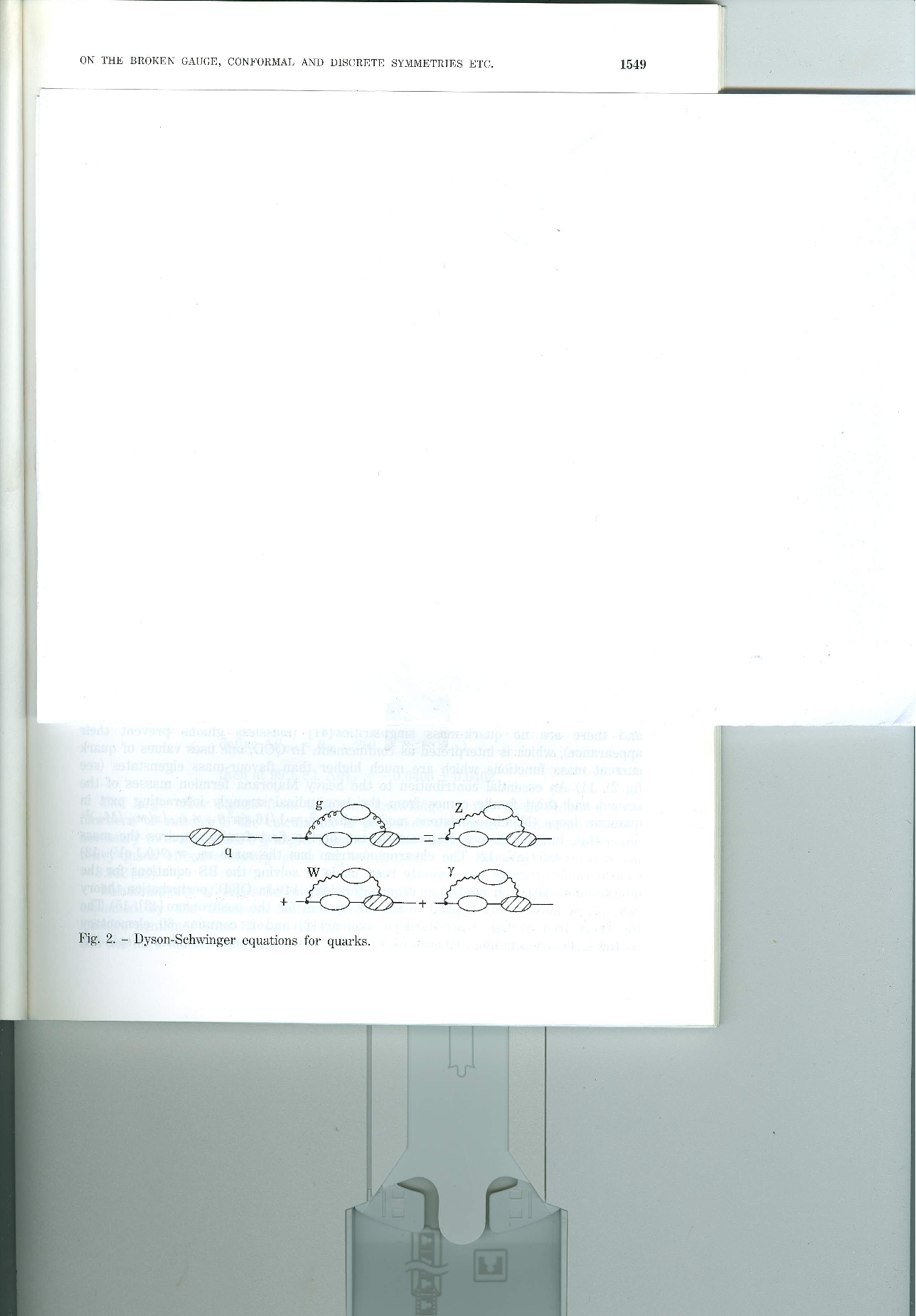}
\end{center}
\end{figure}

\end{document}